\documentstyle[preprint,aps]{revtex}

 \newcommand \be {\begin{equation}}
\newcommand \bea {\begin{eqnarray} \nonumber }
\newcommand \ee {\end{equation}}
\newcommand \eea {\end{eqnarray}}
 
 \newcommand \s {\sigma}

\newcommand \la {\lambda}

\begin{document}
\draft
\preprint{MA/UC3M/06/95}
\title{Glassiness in a model without energy barriers}

\author{Felix Ritort}
\address{Departamento de Matematicas,\\
Universidad Carlos III, Butarque 15\\
Legan{\'e}s 28911, Madrid (Spain)\\
E-Mail: ritort@dulcinea.uc3m.es}
\date{\today}
\maketitle

\begin{abstract}
We propose a microscopic model without energy barriers in order to
explain some generic features observed in structural glasses. The statics can
be exactly solved while the dynamics has been clarified using Monte Carlo
calculations. Although the model has no thermodynamic transition
it captures some of the essential features of real glasses,
i.e., extremely slow relaxation, time dependent hysteresis effects,
anomalous increase of the relaxation time and aging. This suggests that
the effect of entropy barriers can be an important ingredient to account
for the behavior observed in real glasses.
\end{abstract}

\vfill
\pacs{61.20.Lc, 64.70.Pf}

\vfill

\narrowtext
The nature of the glass transition is a longly debated question of much
theoretical interest \cite{Gotze}. It is widely believed that the glass
transition is mainly a dynamical process where the system can remain
trapped in a metastable phase of a finite lifetime depending on the rate
of the cooling process. If glasses are slowly cooled from the
high-temperature region then it is possible to reach a crystal phase of
very low entropy. But if the system is fastly quenched then it reaches a
non-equilibrium regime characterized by the existence of very slow
relaxation phenomena. Usually, the origin of these very slow relaxations
is explained by the existence of a large number of metastable states
separated by energy barriers \cite{Gotze}. The heights of the energy
barriers are widely distributed and the system gets trapped in this
metastable phase during its time evolution.

Recently, there have been developments towards a mean-field theory of
glasses. In those cases one studies systems without quenched disorder
with the aid of replica theory \cite{BoMe}. One finds the existence of a
dynamical transition $T_D$ where the correlation time diverges. Below
that temperature the system is always off equilibrium and relaxes
towards a dynamical state of higher energy. At a lower temperature
replica symmetry breaks and a large number of states dominate the
statics. Below the dynamical transition temperature the system remains
trapped in a very complex free energy landscape with huge free energy
barriers. In mean-field theory, the height of these free energy barriers
increases exponentially fast with $N$ and metastable states have an
infinite lifetime. Generally speaking, free energy barriers get
contributions from an energetic part and an entropic part. In real
structural glasses we expect the effects of energy barriers to be
substantially different from that in mean-field theory because of the
existence of nucleation processes \cite{KiWo}. Nevertheless, the effect
of entropy barriers should be not so dependent on the range of the
interaction. Then we expect entropy barriers to be a relevant mechanism
in mean-field as well as in short-range models.

The purpose of this work is to understand the role of entropy barriers
in the behavior observed in structural glasses. By entropy barriers we
mean the existence of a very small number of directions in phase space
where the energy decreases. We propose a mean-field model with purely
entropy barriers and without metastable states. The phase space of this
model is very simple and resembles to a golf-hole landscape. It mainly
consists of flat directions in energy with a very small number of
channels where the energy decreases.  Although the model has no
thermodynamic transition it shows a behavior reminiscent of real
glasses.


{\em The model.} Let us suppose $N$ distinguishable identical particles
which can occupy $N$ different states. The Hamiltonian is defined by

\be
H=-2\sum_{r=1}^N\,\delta_{N_r\,0}
\label{eq1}
\ee

and the energy per site is given by the fraction of occupied states.
The $N_r$ are the number of particles which occupy the state $r$ (
$r$ runs from 1 to $N$) and they satisfy the constraint
\be
\sum_r\,N_r=N
\label{eq2}
\ee

The model defined in eq.(\ref{eq1}) can be mapped into a Potts model with
a large number of states $N$ with Hamiltonian
\be
H=-\sum_{r=1}^{N}|m_r|
\label{eq3}
\ee

where $m_r$ is the magnetization of state $r$, $m_r=N_r-1$. Looking at
the simple Hamiltonian of eq.(\ref{eq1}) we observe that there is a
trivial ground state with energy per particle
$e_{GS}=-2(1-\frac{1}{N})=-2$ in the large $N$ limit. In this ground
state all the $N$ particles occupy one state, its degeneracy being equal
to $N$. The evolution of the system as a function of the temperature is
as follows. At high temperatures we expect all configurations to have
the same probability and the energy per particle in this limit is
$-2(\frac{N-1}{N})^N=-2/e$. As the temperature is decreased the number
of occupied states decreases while the occupied states increase their
occupation numbers $N_r$. Let us suppose we introduce a dynamics for the
model eq.(\ref{eq1}) (which gives the equilibrium Boltzmann
distribution). The time evolution of the system as the temperature is
decreased is as follows. The rate of variation of the number of
particles in one state increases due to the particles which reach that
state and decreases due to the particles which leave that state. The
energy decreases when one state is emptied during the dynamical
process. Because the total number of particles is conserved, as the
number of occupied states decreases the time the system needs to empty a
further state also increases. In this off-equilibrium situation the
occupation numbers $N_r$ of the occupied states perform a random walk
and the energy eq.(\ref{eq1}) decreases very slowly to the static value
at that temperature. This model has no dynamical phase transition but it
shows the onset of very slow relaxations in the low temperature region
(below $T\simeq 0.2$ close to the maximum of the specific heat).  We
will see that the main characteristics of this model are: strong
dependence of the energy of the system with the cooling rate, hysteresis
effects, anomalous increase of the relaxation time and presence of
aging.

{\em Statics of the model.} In order to solve the statics of this model
we have to compute the partition function. We will suppose that
we have $N$ particles and each particle $i$ is associated with a
variable $\s_i$ which can take $N$ possible values according to the
state the particle occupies.

The partition function is given by,
\be
Z=\frac{1}{N!}\sum_{\s_i}exp(2\beta\sum_{r=1}^N\,\delta_{N_r\,0})
\label{eq4}
\ee

where the factor $N!$ in the denominator is a normalization constant in
order to make the free energy extensive with $N$.  The ocupation numbers
$N_r$ satisfy the constraints $\sum_r N_r=N$ and
$N_r=\sum_{i=1}^N\,\delta_{\s_i\,r}$. Eq.(\ref{eq4}) can be rewritten in
term of occupation numbers as

\be
Z=\sum_{\lbrace N_r=0\rbrace}^N\,\frac{1}{\prod_{r=1}^N\,N_r!}
exp(2\beta\sum_{r=1}^N\,\delta_{N_r\,0})\delta(\sum_r\,N_r-N)
\label{eq5}
\ee

We use the integral representation for the delta function,

\be
\delta(m)=\frac{1}{2\pi}\int_0^{2\pi}d\la e^{i\la m}
\label{eq6}
\ee

where $m$ is an integer. Substitution into eq.(\ref{eq5}) leads to,

\be
Z=\int_{0}^{2\pi}\frac{d\la}{2\pi}expN\Bigl (-i\la+log(\sum_{M=0}^N
\frac{exp(i\la M + 2\beta\delta_{M\,0})}{M!})\Bigr )
\label{eq7}
\ee

The integral in the previous equation
can be readily evaluated by the saddle point method in the large $N$
limit. The saddle point $\la=iz$ gives the free energy $\beta f=-Max_{z}
A(z)$ with

\be
A(z)=-z+log(e^{2\beta}-1+e^{e^z})
\label{eq8}
\ee

The saddle point equation is $e^{2\beta}-1=(y-1)e^y$ where $y=e^z$.
The solution to this equation gives a value $y^*$. The free energy is
given by $f=-y^*/\beta$ and the internal energy $u=
-2e^{2\beta}/(y^*\,e^{y^*})$. We have checked that the first orders in
the high-temperature expansion of eq.(\ref{eq4}) for the energy
coincide with the previous expression. The energy goes to $-2$ at zero
temperature (see fig.1). The specific heat (first derivative of the
energy) increases approximately like $1/T^2$ as the temperature is
decreased and shows a maximum at $T\sim 0.20$. At high temperatures
the entropy converges to $1$ (the number of configurations at infinite
temperature goes like $N^N/N!$, the dominant term for the entropy
being 1). This solution is stable but gives a negative entropy below
$T=1/\beta\simeq 0.345$. Because the entropy at zero temperature
diverges like $log(T)$ (the number of configurations in this limit
being $N/N!$) negative entropies are allowed in the model and there is
no sign of a thermodynamic phase transition.

{\em Dynamics of the model.} We have performed Monte Carlo dynamical
calculations in this model with the Metropolis algorithm.  It is simple
enough to allow very fast computations for very large values of $N$.
We did two kind of numerical experiments. First we performed
annealings starting from a random initial configuration. The
temperature was slowly decreased and the energy was computed over
$t_0$ MC sweep (a MC sweep corresponds to $N$ trials to change the
state of the -randomly chosen- particles). Also we measured the energy
starting from the ground state configuration (only one state occupied)
by slowly increasing the temperature. Results are shown in figure 1
for $N=20000$ and for the static energy. Numerical computations for a
larger number of particles ($N=10^5$) show that finite-size effects
are negligible.  Below $T\simeq 0.17$ we observe a strong dependence
of the energy on the cooling rate and a slow relaxation of the energy
to its equilibrium value. Figure 1 also shows the strong dependence of
the energy on the temperature change rate during the heating
process. The numerical data merges to the static result at a certain
temperature. This is also the temperature at which the energy departs
from the static value in the cooling procedure.  The dependence of
this merging temperature on the time spent on the cooling rate is an
estimate of the relaxation time.

We want now to show that the energy converges to its equilibrium
value. We have studied the relaxation of the energy at zero temperature
starting from an initial random distribution of particles. Because there
are no metastable states the system can reach the ground state. We have
measured the time the system takes to reach the ground state at zero
temperature as a function of the number of particles. We have computed
$\langle log(\tau)\rangle$ for different values of $N$ ranging from $5$
to $20$ (the average $\langle ... \rangle$ means average over different
random initial conditions). We find the typical time very well described
by $\tau\simeq 0.39\,exp(0.67N)$. This means that the system takes an
exponentially large time to reach the ground state. We have not succeded
in deriving an exact expression for the decay of the energy at zero
temperature in the infinite $N$ limit. The problem, being highly non
trivial, can be approximated taking into account the previous result for
the exponentially growing time. We argue that the time $dt$ the system
needs to decrease the fraction of occupied states in a quantity
$d(\frac{N_{oc}}{N})$ scales like

\be
dt\simeq exp(\frac{N}{N_{oc}})d(\frac{N_{oc}}{N})
\label{eqt}
\ee

For finite values of $N$ this expression yields an exponentially large
time for reaching the ground state.  The previous expression means that
for a small number of occupied states the rate of decrease of the energy
is also small (there are less states with more particles per state to be
emptied). Using $u=-2(1-N_{oc}/N)$ we get for the decay of the energy

\be
t=\int_{u_0}^u \,du\,exp\Bigl
(\frac{2}{2+u}\Bigr )
\label{int}
\ee

where $u_0$ is the initial energy at time zero. We have measured the
decay of the energy as a function of time.  While this expression is
only approximate it shows a remarkable agreement with the numerical data
especially in the large time region over several decades of time. The
relaxation of the energy, far from being of a logarithmic or algebraic
type, is extremely slow as eq.(\ref{int}) shows. The fact that the
energy converges to the static result at zero temperature suggests that
a dynamical transition at a finite temperature is absent. In what
follows we will check this point by computing the relaxation time.


{\em The relaxation time and aging.} To fully characterize the dynamics
of this model we have computed the relaxation time. We can define two
types of correlation functions, one for the $\s_i$ variables, the other
one for the energy state variables. In the regime of low temperatures
the system performs a random walk changing particles from one state to
another and the appropiate correlation function is given by the
state to state energy function,

\be
C_e(t,t')=\frac{\sum_{r=1}^N(e_r(t')-u(t'))(e_r(t)-u(t))}
{\sum_{r=1}^N(e_r(t)-u(t))^2}
\label{eq9}
\ee

where $e_r(t)=\delta_{N_r(t)\,0}$ and $u(t)$ is the mean energy per site
at time $t$. We have normalized it in order to have $C_e(t,t)=1$.  For
times $t$ larger than the correlation time $\tau(T)$ the $C_e(t,t')$
should depend only on the time difference $t'-t$ ($t'<t$) and decay
exponentially with time ($C_e(t,t')\sim exp(-(t'-t)/\tau)$). For
$t<<\tau(T)$ the system is off-equilibrium, aging efects are present and
time translation invariance is broken.  We have measured the relaxation
time as a function of the temperature in the low temperature
region. Calculations for this model are fast enough to allow to compute
several orders of magnitude in the relaxation time.  Results are shown
in figure 2. There is no sign of algebraic divergence at any finite
temperature. While a pure Arrhenius divergence $\tau\sim exp(A/T)$ does
not fit enough well (data present a systematic curvature in figure 2) we
find that a Vogel-Fulcher law $\tau\sim exp(A/(T-T_0))$ \cite{VoFu}
describes extremely well the increase of the relaxation time. The value
of $T_0\simeq 0.02$ is stable to including more points in the fit and is
definitely better than the Arrhenius one. We do not attach special
physical meaning to the value of $T_0$ but the indication of an anomaly
in the divergence of the relaxation time. This is different to the
case of models where metastability is present where $T_0$ can be
identified as a thermodynamic transition temperature \cite{Parisi}.

The dynamical transition in this model takes place at $T=0$ and we
expect for $t$ finite the correlation function of eq.(\ref{eq9}) to
display aging.  Introducing the waiting time $t_w$ and redefining the
times $t_w=t,\,\,t'=t+t_w$ in eq.(\ref{eq9}) we find that
$C_e(t_w,t_w+t)$ is pretty well described by the scaling law

\be
C_e(t_w,t_w+t)=f(\frac{t}{t_w})
\label{eq10}
\ee

Similarly to other mean-field models \cite{CuKu,Bou}, it is very
plausible that this scaling behavior is exact in the large $t_w$ limit
at least at zero temperature. Results are shown in figure 3. Data
collapse in a single curve, the scaling function $f$ of eq.(\ref{eq10})
scales like $f(x)\simeq 0.78\,x^{-0.70}$ for large values of $x$. If we
define $u_{EA}=\lim_{t_w\to\infty}C_e(t_w,2t_w)$ we find that this value
jumps discontinuously to a finite value $\simeq 0.58$ at $T=0$ being
zero at any finite $T$ in agreement with the absence of any finite $T$
dynamical transition.

We can summarize now our results. We have introduced a simple model
without energy barriers (i.e. without metastable states) and without
disorder. We have exactly solved the statics while the dynamics has been
clarified by numerical computations of the model. Relaxation is
extremely slow at low temperatures due to the presence of high entropy
barriers, i.e. existence of small number directions in phase space where
energy decreases. While the system has no thermodynamic transition it
displays the main features of real glasses, i.e. extremely slow
relaxation, time dependent hysteresis effects, anomalous increase of the
relaxation time and aging. This suggests that the presence of activated
energy barriers are not the only possible ingredient in order to find a
Vogel-Fulcher behavior (and eventually Arrhenius behavior) for some
transport quantities. The fact that the relaxation time can be nicely
fitted to a Vogel-Fulcher law in this model is an indication that the
existence of anomalies in the relaxation time can be affected by entropy
barrier effects.  The behavior observed in this model suggests that it
is not easy to disentangle the effects of energy barriers form the
effects of entropy barriers, at least experimentally. Real glasses do
have energy barriers and it is clear that this model cannot explain all
the experimentally observed features (for instance, existence of a
cristallization transition). Interestingly enough, it presents the main
dynamical features observed in structural glasses.  The presented model
is of mean-field type because a spatial arrangement of the states is
absent and it is possible to include spatial correlations. Also one can
consider generalizations of this model, for instance, defining the
energy as given by the number of occupied states above a certain finite
level $k$. In this last case we expect to find similar results to those
presented here.

I am grateful to J. P. Bouchaud, M. Mezard , G. Parisi and J. M. Rubi
for very interesting suggestions.  Also I am grateful to L. Bonilla and
A. Wacker for a careful reading of the manuscript. This work was
supported by Universidad Carlos III de Madrid.

\vfill\eject

\vfill
\newpage
{\bf Figure Captions}
\begin{itemize}

\item[Fig.~1] Energy as a function of temperature for two different
processes (cooling and heating) and three different cooling rates
$t_0$ (defined as the number of MC sweeps per temperature step, this
being 0.005). The continuous line is the static energy.

\item[Fig.~2] Relaxation time as a function of $1/T$. The
continuous line is the Vogel-Fulcher law $\tau=A\,exp(B/(T-T_0)$ with
parameters $A=0.89,B=0.51,T_0=0.02$. The dotted line is the fitted
Arrhenius law $\tau=A\,exp(B/T)$ with $A=0.63,B=0.66$. Error bars for
the relaxation time are of the size of the crosses.

\item[Fig.~3] $C_e(t_w,t_w+t)$ for $N=100000$ as a function of $t/t_w$ for
different $t_w$ values at zero temperature.

\end{itemize}


\newpage
\begin{thebibliography}{99}

\bibitem{Gotze} W. G{\"o}tze, {\em ``Liquid, Freezing and Glass
Transition}, Les Houches (1989), J. P. Hansen, D. Levesque, and
J. Zinn-Justin editors, North-Holland

\bibitem{BoMe} J. P. Bouchaud and M. Mezard, {\em J. Phys. I (France)}
{\bf 4}, 1109 (1994)\\ E. Marinari, G. Parisi, and F. Ritort, {\em
J. Phys. A} {\bf 27}, 7615 (1994); {\em J. Phys. A} {\bf 27}, 7647
(1994);

\bibitem{KiWo} For a discussion in the context of spin glasses see
T. R. Kirkpatrick and D. Thirumalai, {\em Phys. Rev. B} {\bf 36}, 5388
(1987); T. R. Kirkpatrick and P. Wolyness, {\em Phys. Rev. B} {\bf
36}, 8552 (1987);

\bibitem{VoFu} H. Vogel, Z. Phys. {\bf 22}, 645 (1921); G. S. Fulcher,
J. Am. Ceram. Soc., {\bf 6}, 339 (1925)

\bibitem{Parisi} G. Parisi, {\em Preprint/cond-mat} {\bf 9411115}

\bibitem{CuKu} L. F. Cugliandolo and J. Kurchan, {\em Phys. Rev. Lett.}
{\bf 71}, 173 (1993); S. Franz and M. Mezard, {\em Europhys. Lett.} {\bf
26}, 209 (1994); E. Marinari and G. Parisi, {\em J. Phys. A} {\bf 26},
L1149 (1993);

\bibitem{Bou} J. P. Bouchaud, {\em J. Phys. I (France)} {\bf 2}, 1705
(1992); J. P. Bouchaud and D. S. Dean, {\em J. Phys. I (France)} {\bf
5}, 265 (1995);

\end{thebibliography}
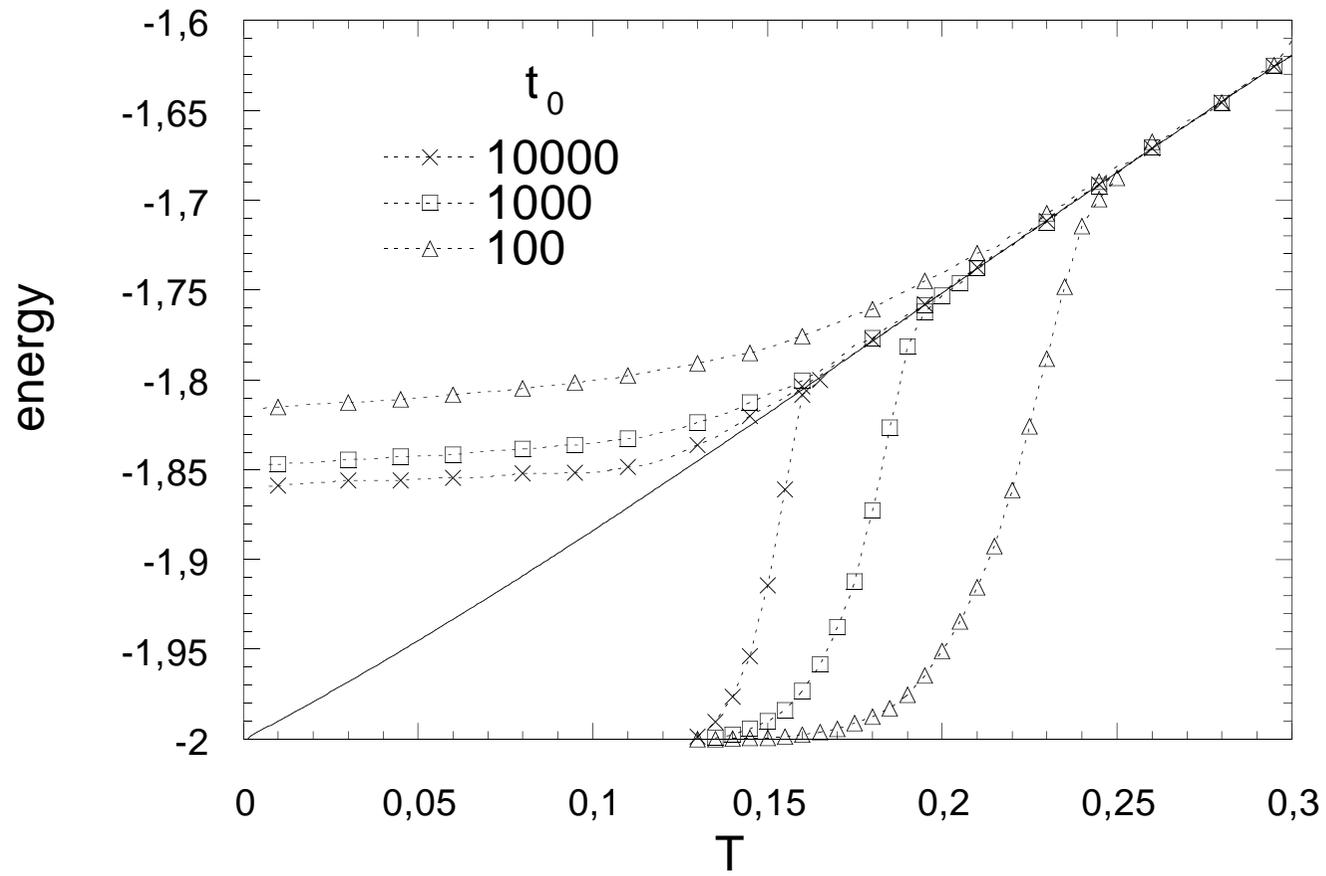
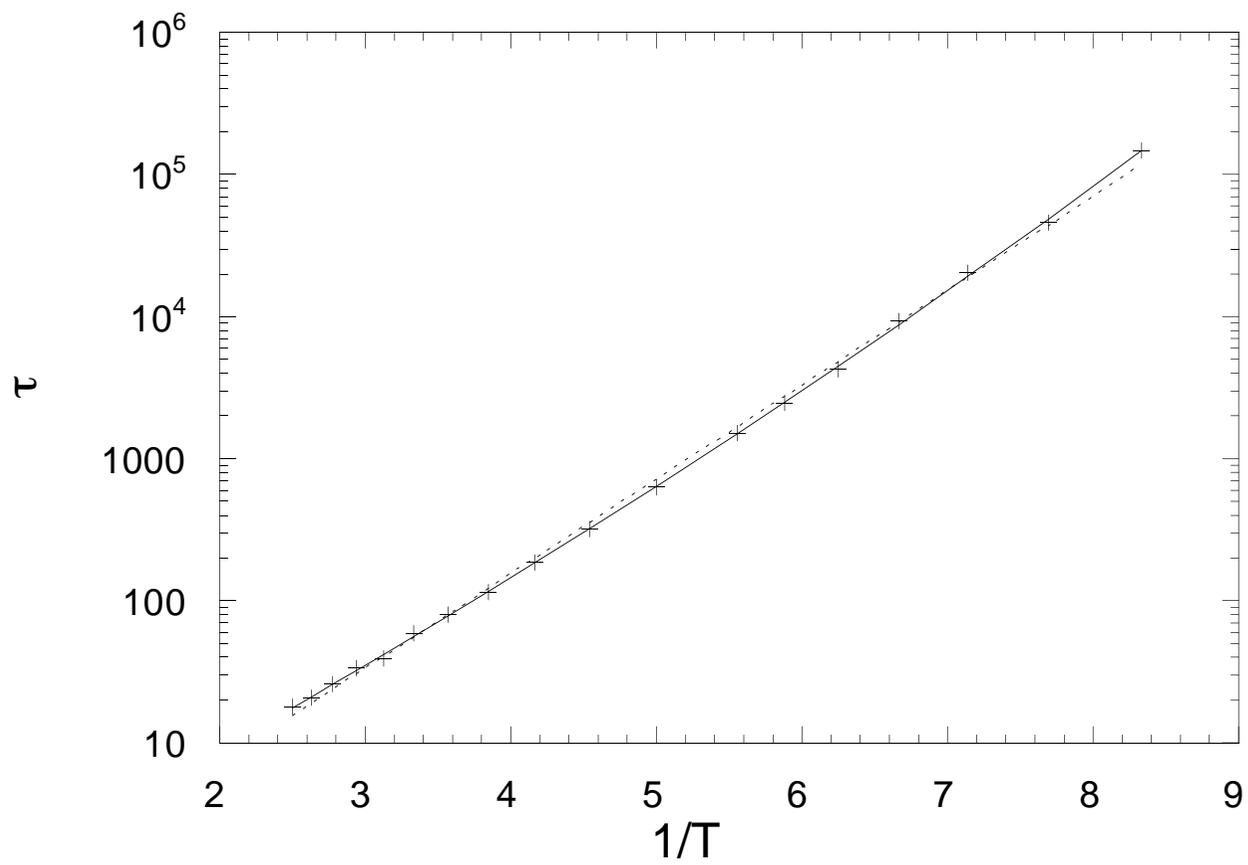
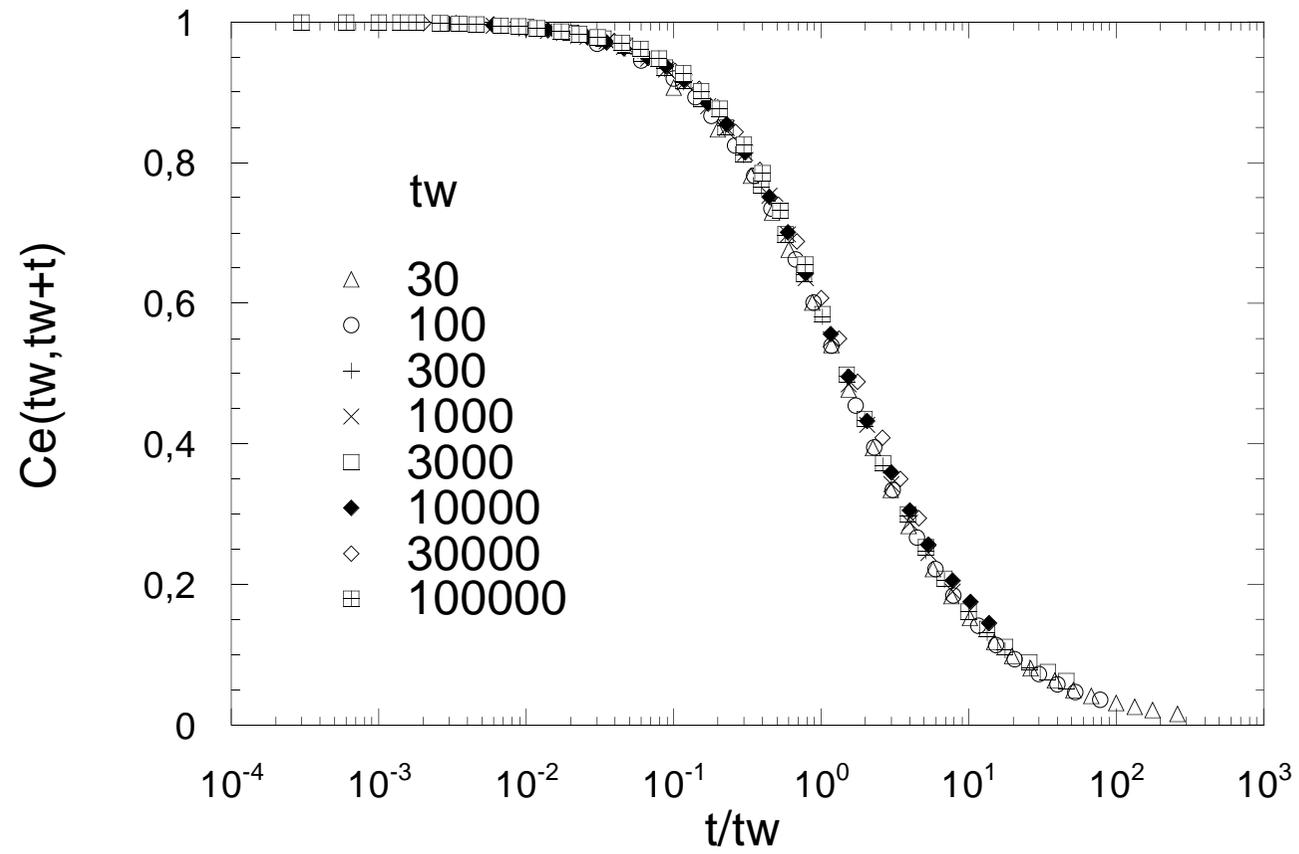
\end{document}